\documentclass[showpacs,aps,graphicx]{revtex4}
\usepackage{graphicx}

\begin{document}
\title{Protecting sing-photon multi-mode W state from photon loss}
\author{Yu-Bo Sheng,$^{1,2}$\footnote{Email address:
shengyb@njupt.edu.cn} Yang Ou-Yang,$^{1,2}$  Lan Zhou,$^{2,3}$  Lei Wang$^{1,2}$}
\address{$^1$ Institute of Signal Processing  Transmission, Nanjing
University of Posts and Telecommunications, Nanjing, 210003,  China\\
 $^2$Key Lab of Broadband Wireless Communication and Sensor Network
 Technology,
 Nanjing University of Posts and Telecommunications, Ministry of
 Education, Nanjing, 210003,
 China\\
 $^3$College of Mathematics \& Physics, Nanjing University of Posts and Telecommunications, Nanjing,
210003, China\\}
\date{\today}

\begin{abstract}
Single-photon entanglement is of major importance in current quantum communications. However, it is sensitive to photon loss.
 In this paper, we
 discuss  the protection of  single-photon multi-mode W state with noiseless linear amplification. It is shown that the amplification factor is only decided with the transmission coefficient of the variable fiber beam splitters, and it does not change with the number of the spatial mode. This protocol may be useful in current quantum information processing.
\end{abstract}
\maketitle

\section{Introduction}
Quantum entanglement is the key element in current quantum communication and quantum computation \cite{book,rmp}. Most quantum communication branches
 require the quantum entanglement. For example, the quantum teleportation \cite{teleportation},  quantum key distribution (QKD) \cite{key1,key2,key3}, and quantum secure direct communication (QSDC) \cite{QSDC1,QSDC2},  all
 need the entanglement to set up the quantum channel. Among all the types of the entanglement, the single-photon entanglement of the form $\frac{1}{\sqrt{2}}(|0\rangle_{A}|1\rangle_{B}+|1\rangle_{A}|0\rangle_{B})$ is the simplest, but it is
 of vice importance \cite{single-particle1,single-particle2,single-particle3}. Here the $|1\rangle$ means one photon, and $A$ and $B$ are the different spatial modes.
 The single-photon entanglement can be created with the 50:50 beam splitter (BS). The single-photon is widely used in the
 quantum repeaters. For example, in the well know the DLCZ protocol, the entanglement of the atomic ensembles of the form $\frac{1}{\sqrt{2}}(|e\rangle_{A}|g\rangle_{B}+|g\rangle_{A}|e\rangle_{B})$ can be converted into the single-photon entanglement \cite{memory,singlephotonrepeater3}.
Here the $|g\rangle$ and $|e\rangle$ are the ground state and excited state, respectively. The entanglement purification and concentration for single-photon
entanglement were also discussed \cite{single-particle4,single-particle5,shengqic,shengoc}.

 On the other hand, if we extend the
single-photon two-mode  entangled state into the multi-mode entangled state of the form of
\begin{eqnarray}
|\Phi\rangle=\frac{1}{\sqrt{N}}(|1\rangle|0\rangle|0\rangle\cdots|0\rangle+|0\rangle|1\rangle|0\rangle\cdots|0\rangle+|0\rangle|0\rangle\cdots|0\rangle|1\rangle),
\end{eqnarray}
it is the single-photon multi-mode W state \cite{nonlocality,telescope,concentration}. Heaney \emph{et al.} discussed the nonlocality of a system containing a single photon W state \cite{nonlocality}.
Especially, Gottesman \emph{et al.} showed that such single-photon W state is extremely useful in longer-baseline telescopes \cite{telescope}.

However, like any other quantum entanglement based on a two-photon polarized entangled
pair or two-mode electromagnetic field, single-photon
entanglement is also sensitive to decoherence. The noise will make the photon loss. One of the powerful ways to overcome the photon loss
is the photon noiseless linear amplification (NLA) \cite{amplification1,amplification2,amplification3,amplification4,amplification5,amplification6,amplification7,amplification8,amplification9,amplification10,amplification11}, which was first proposed by Ralph and Lund \cite{amplification1}. The way of heralded qubit amplifier was
discussed for device-independent QKD for several groups \cite{amplification2,amplification3,amplification4}.   Xiang \emph{et al.} realized the  heralded NLA and distillation of
entanglement \cite{amplification5,amplification6}.  Osorio \emph{et al.} also reported their experiment result for heralded photon amplification for quantum communication \cite{amplification7}.
Recently, Zhang \emph{et al.} developed the NLA to protect the two-mode single-photon entanglement from photon loss \cite{amplification8}.
Though several groups have discussed the qubit amplification, their protocols are all based on the loss of single photon, or the two-mode single photon entanglement. None protocol discuss the qubit amplification for the multi-mode single-photon W state.

Therefore, in this paper, we will present a linear photon amplification protocol for multi-mode single-photon W state. It is shown that the multi-mode single-photon W state can also be protected with the help of local single photons.

\section{Protecting the single-photon multi-mode W entangled state}
 As shown in Fig. 1,
suppose that three parties A, B, and C share an initial state with the form of
\begin{eqnarray}
|\Phi\rangle_{ABC}=\frac{1}{\sqrt{3}}(|1\rangle_{a1}|0\rangle_{a2}|0\rangle_{a3}+|0\rangle_{a1}|1\rangle_{a2}|0\rangle_{a3}+|0\rangle_{a1}|0\rangle_{a2}|1\rangle_{a3}).
\end{eqnarray}
in the spatial modes a1, a2, and a3.

The noisy channel will make the photon loss with the probability of $1-\eta$. Therefore, the whole state will degrade to
\begin{eqnarray}
\rho_{ABC}=\eta|\Phi\rangle_{ABC}\langle\Phi|+(1-\eta)|vac\rangle\langle vac|.\label{mix1}
\end{eqnarray}

\begin{figure}[!h]
\begin{center}
\includegraphics[width=8cm,angle=0]{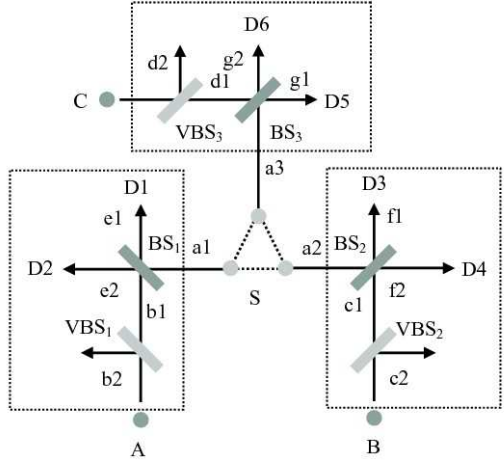}
\caption{A schematic drawing of our linear optical amplification protocol for the three-mode single-photon W state. Due to the photon loss, the three parties A, B, and C share a mixed state as $\rho_{ABC}=\eta|\Phi\rangle_{ABC}\langle\Phi|+(1-\eta)|vac\rangle\langle vac|$. In the protocol, each of the three parties needs to prepare a local single photon and make them pass through the VBSs, respectively. With the help of the VBSs and BSs, we can distill a new mixed state as $\rho_{ABC}=\eta'|\Phi\rangle_{ABC}\langle\Phi|+(1-\eta')|vac\rangle\langle vac|$. If the transmission of the VBSs meets $t<\frac{1}{2}$, we can obtain $G=\frac{\eta'}{\eta}>1$.}
\end{center}
\end{figure}
The aim for  protecting  the single-photon entanglement is essentially to increase the probability $\eta$. As shown in Fig. 1, three variable fiber beam splitters (VBSs) are used in our protocol. The transmission of the three VBSs are the same, which is called t. In order to increase the probability of the W state, the three parties  prepare a local single photon, respectively. The three
single photons named $|1\rangle_{A}$, $|1\rangle_{B}$ and $|1\rangle_{C}$ are sent into the VBSs, respectively. The VBSs will produce the entangled state between different spatial modes as
\begin{eqnarray}
|1\rangle_{A}\rightarrow\sqrt{t}|1\rangle_{b1}|0\rangle_{b2}+\sqrt{1-t}|0\rangle_{b1}|1\rangle_{b2},\nonumber\\
|1\rangle_{B}\rightarrow\sqrt{t}|1\rangle_{c1}|0\rangle_{c2}+\sqrt{1-t}|0\rangle_{c1}|1\rangle_{c2},\nonumber\\
|1\rangle_{C}\rightarrow\sqrt{t}|1\rangle_{d1}|0\rangle_{d2}+\sqrt{1-t}|0\rangle_{d1}|1\rangle_{d2}.
\end{eqnarray}
\begin{figure}[!h]
\begin{center}
\includegraphics[width=8cm,angle=0]{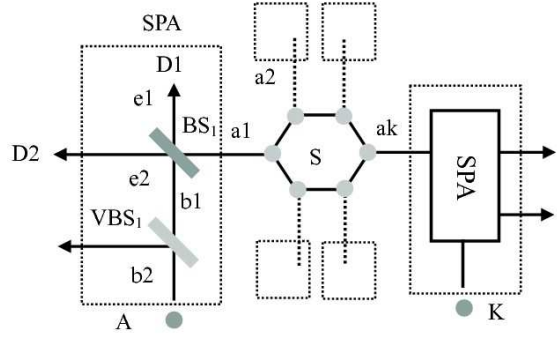}
\caption{Our amplification protocol can be extended to deal with the multi-mode single-photon W state. Suppose the N parties share a mixed state as
$\rho_{N}=\eta_{N}|\Phi\rangle_{N}\langle\Phi|+(1-\eta_{N})|vac\rangle\langle vac|$. Similarly, each of the N parties needs to prepare a local single photon, and makes them pass through N local VBSs, respectively. With the help of the VBSs and BSs, we can still distill a new mixed state as $\rho_{N}'=\eta_{N}'|\Phi\rangle_{N}\langle\Phi|+(1-\eta_{N}')|vac\rangle\langle vac|$. Interestingly, we can also obtain the amplification factor $G>1$ under the case that $t<\frac{1}{2}$.}
\end{center}
\end{figure}
 Therefore, the whole system is in the state $|\Phi\rangle_{ABC}\otimes|1\rangle_{A}\otimes|1\rangle_{B}\otimes|1\rangle_{C}$ with the probability of $\eta$, or in the state  $|vac\rangle\otimes|1\rangle_{A}\otimes|1\rangle_{B}\otimes|1\rangle_{C}$ with the probability of $1-\eta$.
We first discuss the first case. After the VBSs, it will evolve as
\begin{eqnarray}
&&|\Phi\rangle_{ABC}\otimes|1\rangle_{A}\otimes|1\rangle_{B}\otimes|1\rangle_{C}\nonumber\\
&=&\frac{1}{\sqrt{3}}(|1\rangle_{a1}|0\rangle_{a2}|0\rangle_{a3}+|0\rangle_{a1}|1\rangle_{a2}|0\rangle_{a3}+|0\rangle_{a1}|0\rangle_{a2}|1\rangle_{a3})
\otimes|1\rangle_{A}\otimes|1\rangle_{B}\otimes|1\rangle_{C}\nonumber\\
&\rightarrow&\frac{1}{\sqrt{3}}(|1\rangle_{a1}|0\rangle_{a2}|0\rangle_{a3}+|0\rangle_{a1}|1\rangle_{a2}|0\rangle_{a3}+|0\rangle_{a1}|0\rangle_{a2}|1\rangle_{a3})\nonumber\\
&\otimes&(\sqrt{t}|1\rangle_{b1}|0\rangle_{b2}+\sqrt{1-t}|0\rangle_{b1}|1\rangle_{b2})\otimes(\sqrt{t}|1\rangle_{c1}|0\rangle_{c2}+\sqrt{1-t}|0\rangle_{c1}|1\rangle_{c2})\nonumber\\
&\otimes&(\sqrt{t}|1\rangle_{d1}|0\rangle_{d2}+\sqrt{1-t}|0\rangle_{d1}|1\rangle_{d2})\nonumber\\
&=&\frac{1}{\sqrt{3}}[(|1\rangle_{a1}|0\rangle_{a2}|0\rangle_{a3}\otimes\sqrt{t}|1\rangle_{b1}|0\rangle_{b2})\otimes(\sqrt{t}|1\rangle_{c1}|0\rangle_{c2}+\sqrt{1-t}|0\rangle_{c1}|1\rangle_{c2})\nonumber\\
&\otimes&(\sqrt{t}|1\rangle_{d1}|0\rangle_{d2}+\sqrt{1-t}|0\rangle_{d1}|1\rangle_{d2})\nonumber\\
&+&(|1\rangle_{a1}|0\rangle_{a2}|0\rangle_{a3}\otimes\sqrt{1-t}|0\rangle_{b1}|1\rangle_{b2})\otimes(\sqrt{t}|1\rangle_{c1}|0\rangle_{c2}+\sqrt{1-t}|0\rangle_{c1}|1\rangle_{c2})\nonumber\\
&\otimes&(\sqrt{t}|1\rangle_{d1}|0\rangle_{d2}+\sqrt{1-t}|0\rangle_{d1}|1\rangle_{d2})\nonumber\\
&+&(|0\rangle_{a1}|1\rangle_{a2}|0\rangle_{a3}\otimes\sqrt{t}|1\rangle_{c1}|0\rangle_{c2})\otimes(\sqrt{t}|1\rangle_{b1}|0\rangle_{b2}+\sqrt{1-t}|0\rangle_{b1}|1\rangle_{b2})\nonumber\\
&\otimes&(\sqrt{t}|1\rangle_{d1}|0\rangle_{d2}+\sqrt{1-t}|0\rangle_{d1}|1\rangle_{d2})\nonumber\\
&+&(|0\rangle_{a1}|1\rangle_{a2}|0\rangle_{a3}\otimes\sqrt{1-t}|0\rangle_{c1}|1\rangle_{c2})\otimes(\sqrt{1-t}|0\rangle_{b1}|1\rangle_{b2}+\sqrt{1-t}|0\rangle_{b1}|1\rangle_{b2})\nonumber\\
&\otimes&(\sqrt{t}|1\rangle_{d1}|0\rangle_{d2}+\sqrt{1-t}|0\rangle_{d1}|1\rangle_{d2})\nonumber\\
&+&(|0\rangle_{a1}|0\rangle_{a2}|1\rangle_{a3}\otimes\sqrt{t}|1\rangle_{d1}|0\rangle_{d2})\otimes
(\sqrt{t}|1\rangle_{b1}|0\rangle_{b2}+\sqrt{1-t}|0\rangle_{b1}|1\rangle_{b2})\nonumber\\
&\otimes&(\sqrt{t}|1\rangle_{c1}|0\rangle_{c2}+\sqrt{1-t}|0\rangle_{c1}|1\rangle_{c2})\nonumber\\
&+&(|0\rangle_{a1}|0\rangle_{a2}|1\rangle_{a3}\otimes\sqrt{1-t}|0\rangle_{d1}|1\rangle_{d2})\otimes
(\sqrt{1-t}|0\rangle_{b1}|1\rangle_{b2}+\sqrt{1-t}|0\rangle_{b1}|1\rangle_{b2})\nonumber\\
&\otimes&(\sqrt{t}|1\rangle_{c1}|0\rangle_{c2}+\sqrt{1-t}|0\rangle_{c1}|1\rangle_{c2})].\label{elove1}
\end{eqnarray}
On the the other hand, the other case $|vac\rangle\otimes|1\rangle_{A}\otimes|1\rangle_{B}\otimes|1\rangle_{C}$  will evolve as
\begin{eqnarray}
&&|vac\rangle\otimes|1\rangle_{A}\otimes|1\rangle_{B}\otimes|1\rangle_{C}\rightarrow(\sqrt{t}|1\rangle_{b1}|0\rangle_{b2}+\sqrt{1-t}|0\rangle_{b1}|1\rangle_{b2})\nonumber\\
&\otimes&(\sqrt{t}|1\rangle_{c1}|0\rangle_{c2}+\sqrt{1-t}|0\rangle_{c1}|1\rangle_{c2})\otimes(\sqrt{t}|1\rangle_{d1}|0\rangle_{d2}+\sqrt{1-t}|0\rangle_{d1}|1\rangle_{d2})\nonumber\\
&=&\sqrt{t^{3}}|1\rangle_{b1}|0\rangle_{b2}|1\rangle_{c1}|0\rangle_{c2}|1\rangle_{d1}|0\rangle_{d2}+\sqrt{(1-t)t^{2}}|1\rangle_{b1}|0\rangle_{b2}|1\rangle_{c1}|0\rangle_{c2}|0\rangle_{d1}|1\rangle_{d2}\nonumber\\
&+&\sqrt{(1-t)t^{2}}|1\rangle_{b1}|0\rangle_{b2}|0\rangle_{c1}|1\rangle_{c2}|1\rangle_{d1}|0\rangle_{d2}+\sqrt{(1-t)^{2}t}|1\rangle_{b1}|0\rangle_{b2}|0\rangle_{c1}|1\rangle_{c2}|0\rangle_{d1}|1\rangle_{d2}\nonumber\\
&+&\sqrt{(1-t)t^{2}}|0\rangle_{b1}|1\rangle_{b2}|1\rangle_{c1}|0\rangle_{c2}|1\rangle_{d1}|0\rangle_{d2}+\sqrt{(1-t)^{2}t}|0\rangle_{b1}|1\rangle_{b2}|1\rangle_{c1}|0\rangle_{c2}|0\rangle_{d1}|1\rangle_{d2}\nonumber\\
&+&\sqrt{(1-t)^{2}t}|0\rangle_{b1}|1\rangle_{b2}|0\rangle_{c1}|1\rangle_{c2}|1\rangle_{d1}|0\rangle_{d2}+\sqrt{(1-t)^{3}}|0\rangle_{b1}|1\rangle_{b2}|0\rangle_{c1}|1\rangle_{c2}|0\rangle_{d1}|1\rangle_{d2}.\nonumber\\\label{elove2}
\end{eqnarray}
 In Fig. 1, the beam splitters (BSs) will divide the photons into different spatial modes. For example, the BS$_{1}$ will make $|1\rangle_{b1}\rightarrow\frac{1}{\sqrt{2}}(|1\rangle_{e1}|0\rangle_{e2}+|0\rangle_{e1}|1\rangle_{e2})$, and $|1\rangle_{a1}\rightarrow\frac{1}{\sqrt{2}}(|1\rangle_{e1}|0\rangle_{e2}-|0\rangle_{e1}|1\rangle_{e2})$.
If both the spatial modes b$_{1}$ and a$_{1}$ contain one photon, after passing through the  BS$_{1}$, the two photons will be both in the $e1$ modes or $e2$ modes. It is also called the Hong-Ou-Mandel bunching effect.
From Eq. (\ref{elove1}), we choose the cases that each output mode of the BSs only contains one photon, which means that the single-photon detectors
D1 or D2, D3 or D4, and D5 or D6 only register one photon. In this way, we can obtain
\begin{eqnarray}
|\Phi\rangle_{1}&=&\sqrt{\frac{t^{2}(1-t)}{3}}(|1\rangle_{b2}|0\rangle_{c2}|0\rangle_{d2}+|0\rangle_{b2}|1\rangle_{c2}|0\rangle_{d2}
+|0\rangle_{b2}|0\rangle_{c2}|1\rangle_{d2}),
\end{eqnarray}
with the probability of $t^{2}(1-t)$. On the other hand, from Eq. ({\ref{elove2}}), the item $\sqrt{t^{3}}|1\rangle_{b1}|0\rangle_{b2}|1\rangle_{c1}|0\rangle_{c2}|1\rangle_{d1}|0\rangle_{d2}$ can also lead the success case and becomes
a vacuum state after the three photons being detected.

After choosing the success case, we will finally obtain a new mixed state as
\begin{eqnarray}
\rho'_{ABC}=\eta'|\Phi\rangle_{ABC}\langle\Phi|+(1-\eta')|vac\rangle\langle vac|.\label{mix2}
\end{eqnarray}
Here $\eta'=\frac{\eta t^{2}(1-t)}{\eta t^{2}(1-t)+(1-\eta)t^{3}}$, and the success probability is $P=\eta t^{2}(1-t)+(1-\eta)t^{3}$.
The amplification factor $G$ can be written as
 \begin{eqnarray}
 G\equiv\frac{\eta'}{\eta}=\frac{t^{2}(1-t)}{\eta t^{2}(1-t)+(1-\eta)t^{3}}=\frac{1-t}{\eta(1-t)+(1-\eta)t}.
 \end{eqnarray}
If we require $G>1$, it means that $\frac{\eta'}{\eta}=\frac{t^{2}(1-t)}{\eta t^{2}(1-t)+(1-\eta)t^{3}}>1$. We can easily
obtain $t<\frac{1}{2}$.

So far, the amplification process is completed. Interestingly, we can extend the amplification of single-photon three-mode W state to the  single-photon multi-mode
W state. The maximally entangled single-photon multi-mode W state can be written as
 \begin{eqnarray}
|\Phi\rangle_{N}&=&\frac{1}{\sqrt{N}}(|1\rangle_{a1}|0\rangle_{a2}|0\rangle_{a3}\cdots|0\rangle_{aN}\nonumber\\
&+&|0\rangle_{a1}|1\rangle_{a2}|0\rangle_{a3}\cdots|0\rangle_{aN}+\cdots+|0\rangle_{a1}|0\rangle_{a2}|0\rangle_{a3}\cdots|1\rangle_{aN}).
\end{eqnarray}
If it suffers from the photon loss, it will degrade to a mixed state of the form
\begin{eqnarray}
\rho_{N}=\eta_{N}|\Phi\rangle_{N}\langle\Phi|+(1-\eta_{N})|vac\rangle\langle vac|.\label{mix2}
\end{eqnarray}
\begin{figure}[!h]
\begin{center}
\includegraphics[width=8cm,angle=0]{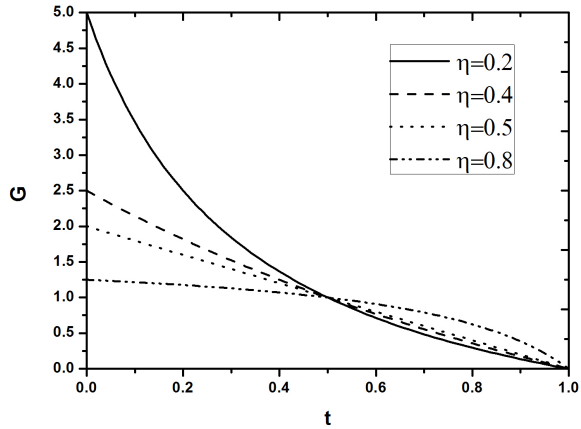}
\caption{The amplification factor G is altered with the transmission coefficient $t$. In order to realize the amplification, we should let $t<\frac{1}{2}$.}
\end{center}
\end{figure}

We can also use the similar setup to perform the amplification. As shown in Fig. 2, we require $N$ VBSs and $N$ single photons to
complete the amplification. With the same principle, the states are in the
$|\Phi\rangle_{N}\otimes|1\rangle_{a1}\otimes|1\rangle_{a2}\otimes\cdots\otimes|1\rangle_{aN}$ with the probability of $\eta$, and in the $|vac\rangle\otimes|1\rangle_{a1}\otimes|1\rangle_{a2}\otimes\cdots\otimes|1\rangle_{aN}$ with the probability of
$1-\eta$. After the VBSs, the item $|\Phi\rangle_{N}\otimes|1\rangle_{a1}\otimes|1\rangle_{a2}\otimes\cdots\otimes|1\rangle_{aN}$ will evolve to
\begin{eqnarray}
&&|\Phi\rangle_{N}\otimes|1\rangle_{a1}\otimes|1\rangle_{a2}\otimes\cdots\otimes|1\rangle_{aN}=
\frac{1}{\sqrt{N}}(|1\rangle_{a1}|0\rangle_{a2}|0\rangle_{a3}\cdots|0\rangle_{aN}\nonumber\\
&+&|0\rangle_{a1}|1\rangle_{a2}|0\rangle_{a3}\cdots|0\rangle_{aN}+\cdots+|0\rangle_{a1}|0\rangle_{a2}|0\rangle_{a3}\cdots|1\rangle_{aN})\nonumber\\
&\otimes&|1\rangle_{a2}\otimes|1\rangle_{a2}\otimes\cdots\otimes|1\rangle_{aN}
\rightarrow\frac{1}{\sqrt{N}}(|1\rangle_{a1}|0\rangle_{a2}|0\rangle_{a3}\cdots|0\rangle_{aN}\nonumber\\
&+&|0\rangle_{a1}|1\rangle_{a2}|0\rangle_{a3}\cdots|0\rangle_{aN}+\cdots+|0\rangle_{a1}|0\rangle_{a2}|0\rangle_{a3}\cdots|1\rangle_{aN})\nonumber\\
&\otimes&(\sqrt{t}|1\rangle_{b1}|0\rangle_{b2}+\sqrt{1-t}|0\rangle_{b1}|1\rangle_{b2})
\otimes(\sqrt{t}|1\rangle_{c1}|0\rangle_{c2}+\sqrt{1-t}|0\rangle_{c1}|1\rangle_{c2})\otimes\cdots\nonumber\\
&\otimes&(\sqrt{t}|1\rangle_{n1}|0\rangle_{n2}+\sqrt{1-t}|0\rangle_{n1}|1\rangle_{n2})\nonumber\\
&=&\frac{1}{\sqrt{N}}(|1\rangle_{a1}|0\rangle_{a2}|0\rangle_{a3}\cdots|0\rangle_{aN}\otimes\sqrt{t}|1\rangle_{b1}|0\rangle_{b2})
\otimes(\sqrt{t}|1\rangle_{c1}|0\rangle_{c2}+\sqrt{1-t}|0\rangle_{c1}|1\rangle_{c2})\nonumber\\
&\otimes&\cdots\otimes(\sqrt{t}|1\rangle_{n1}|0\rangle_{n2}+\sqrt{1-t}|0\rangle_{n1}|1\rangle_{n2})\nonumber\\
&+&|1\rangle_{a1}|0\rangle_{a2}|0\rangle_{a3}\cdots|0\rangle_{aN}\otimes\sqrt{1-t}|0\rangle_{b1}|1\rangle_{b2}
\otimes(\sqrt{t}|1\rangle_{c1}|0\rangle_{c2}+\sqrt{1-t}|0\rangle_{c1}|1\rangle_{c2})\nonumber\\
&\otimes&\cdots\otimes(\sqrt{t}|1\rangle_{n1}|0\rangle_{n2}+\sqrt{1-t}|0\rangle_{n1}|1\rangle_{n2})\nonumber\\
&+&\cdots+|0\rangle_{a1}|0\rangle_{a2}|0\rangle_{a3}\cdots|1\rangle_{aN}\otimes\sqrt{t}|1\rangle_{n1}|0\rangle_{n2}\nonumber\\
&\otimes&(\sqrt{t}|1\rangle_{b1}|0\rangle_{b2}+\sqrt{1-t}|0\rangle_{b1}|1\rangle_{b2})
\otimes(\sqrt{t}|1\rangle_{c1}|0\rangle_{c2}+\sqrt{1-t}|0\rangle_{c1}|1\rangle_{c2})\otimes\cdots\nonumber\\
&\otimes&(\sqrt{t}|1\rangle_{k1}|0\rangle_{k2}+\sqrt{1-t}|0\rangle_{k1}|1\rangle_{k2})\nonumber\\
&+&|0\rangle_{a1}|0\rangle_{a2}|0\rangle_{a3}\cdots|1\rangle_{aN}\otimes\sqrt{1-t}|0\rangle_{n1}|1\rangle_{n2}\nonumber\\
&\otimes&(\sqrt{t}|1\rangle_{b1}|0\rangle_{b2}+\sqrt{1-t}|0\rangle_{b1}|1\rangle_{b2})
\otimes(\sqrt{t}|1\rangle_{c1}|0\rangle_{c2}+\sqrt{1-t}|0\rangle_{c1}|1\rangle_{c2})\otimes\cdots\nonumber\\
&\otimes&(\sqrt{t}|1\rangle_{k1}|0\rangle_{k2}+\sqrt{1-t}|0\rangle_{k1}|1\rangle_{k2}).\label{elove3}
\end{eqnarray}

\begin{figure}[!h]
\begin{center}
\includegraphics[width=8cm,angle=0]{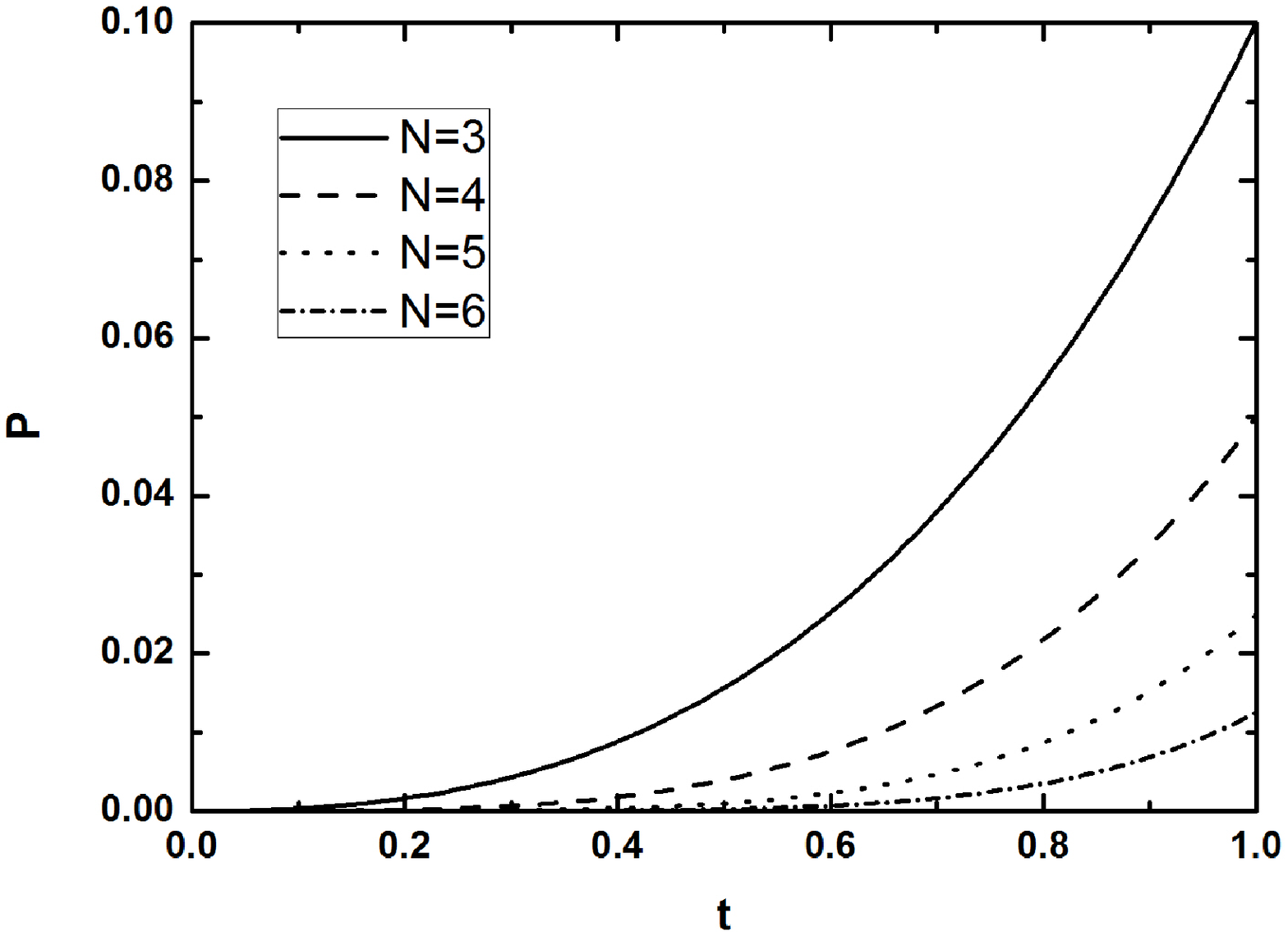}
\caption{The success probability $P$ is altered with the transmission coefficient $t$. We choose $\eta=0.2$.}
\end{center}
\end{figure}

Similarly, the item $|vac\rangle\otimes|1\rangle_{a1}\otimes|1\rangle_{a2}\otimes\cdots\otimes|1\rangle_{aN}$ will evolve to
\begin{eqnarray}
&&|vac\rangle\otimes|1\rangle_{a1}\otimes|1\rangle_{a2}\otimes\cdots\otimes|1\rangle_{aN}\rightarrow(\sqrt{t}|1\rangle_{b1}|0\rangle_{b2}+\sqrt{1-t}|0\rangle_{b1}|1\rangle_{b2})\nonumber\\
&\otimes&(\sqrt{t}|1\rangle_{c1}|0\rangle_{c2}+\sqrt{1-t}|0\rangle_{c1}|1\rangle_{c2})\otimes\cdots
\otimes(\sqrt{t}|1\rangle_{n1}|0\rangle_{n2}+\sqrt{1-t}|0\rangle_{n1}|1\rangle_{n2})\nonumber\\
&=&\sqrt{t^{N}}|1\rangle_{b1}|0\rangle_{b2}|1\rangle_{c1}|0\rangle_{c2}\cdots|1\rangle_{n1}|0\rangle_{n2}\nonumber\\
&+&\sqrt{t^{N-1}(1-t)}(|1\rangle_{b1}|0\rangle_{b2}|0\rangle_{c1}|1\rangle_{c2}\cdots|1\rangle_{n1}|0\rangle_{n2}\nonumber\\
&+&|1\rangle_{b1}|0\rangle_{b2}|1\rangle_{c1}|0\rangle_{c2}|0\rangle_{c1}|1\rangle_{c2}\cdots|1\rangle_{n1}|0\rangle_{n2})\nonumber\\
&+&\cdots+\sqrt{(1-t)^{N}}|0\rangle_{b1}|1\rangle_{b2}|0\rangle_{c1}|1\rangle_{c2}\cdots|0\rangle_{n1}|1\rangle_{n2}.\label{elove4}
\end{eqnarray}
\begin{figure}[!h]
\begin{center}
\includegraphics[width=8cm,angle=0]{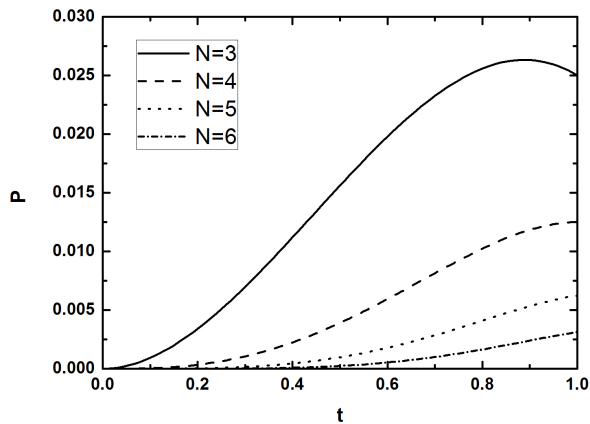}
\caption{The success probability $P$ is altered with the transmission coefficient $t$. We choose $\eta=0.8$.}
\end{center}
\end{figure}

Then, we select the case that it makes one of the single photon detectors after each BS  only detect one photon. The success cases must ensure that only one photon pass through each BS. Therefore, Eq. (\ref{elove3}) will collapse to
 \begin{eqnarray}
 |\Phi\rangle_{N}&=&\sqrt{\frac{(1-t)t^{N-1}}{N}}(|1\rangle_{a1}|0\rangle_{a2}|0\rangle_{a3}\cdots|0\rangle_{aN}
+|0\rangle_{a1}|1\rangle_{a2}|0\rangle_{a3}\cdots|0\rangle_{aN}\nonumber\\
&+&\cdots+|0\rangle_{a1}|0\rangle_{a2}|0\rangle_{a3}\cdots|1\rangle_{aN}),
\end{eqnarray}
with the success probability of $(1-t)t^{N-1}$.

Similarly, in Eq. (\ref{elove4}), we can find that the item $\sqrt{t^{N}}|1\rangle_{b1}|0\rangle_{b2}|1\rangle_{c1}|0\rangle_{c2}\cdots|1\rangle_{n1}|0\rangle_{n2}$ can also lead the success case and finally evolve to the vacuum state after the photon detection. In this way, after the amplification, we can also obtain a mixed state as
 \begin{eqnarray}
\rho_{N}'=\eta_{N}'|\Phi\rangle_{N}\langle\Phi|+(1-\eta_{N}')|vac\rangle\langle vac|,
\end{eqnarray}
where $\eta_{N}'=\frac{\eta(1-t)t^{N-1}}{\eta(1-t)t^{N-1}+(1-\eta)t^{N}}$, and the success probability $P_{N}=\eta_{N}(1-t)t^{N-1}+(1-\eta_{N})t^{N}$. In order to realize the amplification, we must make the amplification factor
 \begin{eqnarray}
G_{N}\equiv\frac{\eta_{N}'}{\eta_{N}}=\frac{(1-t)t^{N-1}}{\eta(1-t)t^{N-1}+(1-\eta)t^{N}}=\frac{1-t}{\eta(1-t)+(1-\eta)t}=G.\label{GN}
\end{eqnarray}
It is shown that the amplification factor $G$ does not change with the number of spatial mode $N$.

\section{Discussion}
In the paper, we put forward an efficient linear noiseless amplification protocol for the multi-mode single-photon W state. In the protocol,  a pair of $N$-mode single-photon W state is shared by N parties. Because of the photon loss, the single-photon W state is degraded to a mixed state as
$\rho_{N}=\eta_{N}|\Phi\rangle_{N}\langle\Phi|+(1-\eta_{N})|vac\rangle\langle vac|$. In order to realize the noiseless amplification,  each party needs to prepare a local single photon, and make it pass through a VBS with the transmission of $t$. Subsequently, each party makes the  photons in his or her mode enter the BS. By selecting the cases that each output mode of the BS only contains one photon, they can distill a new mixed state as
$\rho_{N}'=\eta_{N}'|\Phi\rangle_{N}\langle\Phi|+(1-\eta_{N}')|vac\rangle\langle vac|$. Under the case that the $t<\frac{1}{2}$, we can make the amplification factor $G=\frac{\eta_{N}'}{\eta_{N}}>1$.
Our protocol has some obvious advantages. First, we only require one pair of single-photon W state. 
As the entanglement source is precious, our protocol is economical. Second, we only require the linear optical elements, 
which makes our protocol can be easily realized in current experimental conditions.

VBS is the key element of our protocol. In our protocol, we must require the transmission of each VBS meets $t<\frac{1}2{}$. Actually, VBS is a common linear  optical element in current technology. Recently, Osorio \emph{et al.} also reported their results about heralded photon amplification for quantum communication with the help of the VBS \cite{amplification7}.
   In their experiment, they successfully adjust the splitting ratio of VBS from 50:50 to 90:10 to increase
   the visibility from 46.7 $\pm$ 3.1\% to 96.3 $\pm$ 3.8\%. Based on their experiment, our requirement for $t<\frac{1}2{}$ can be easily obtained.
\begin{figure}[!h]
\begin{center}
\includegraphics[width=8cm,angle=0]{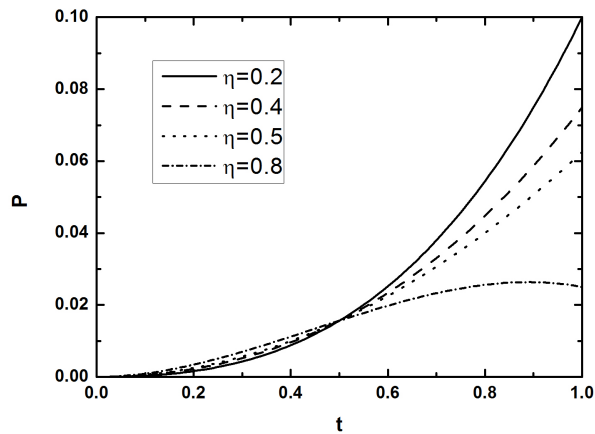}
\caption{The success probability $P$ is altered with the transmission coefficient $t$. We choose $N=3$, and $\eta=0.2,$ 04, 0.6, and 0.8, respectively.}
\end{center}
\end{figure}
In Fig. 3, Fig. 4 and Fig. 5, we calculate the values of amplification factor $G$ and success probability $P$ as a function of the transmission of the VBS (t). In Fig. 3, we fix $N=3$ and calculate the amplification factor $G$ altered with the $t$. Form Fig. 3, if $t\rightarrow0$, $G\rightarrow\frac{1}{\eta}$. In Eq. (\ref{GN}), we can easily  obtain the limitation of $G$ as $\frac{1}{\eta}$. In Fig. 4, we fix  $\eta=0.2$, and the four curves represent the mode number $N$=3, 4, 5, 6, respectively.  In Fig. 5, we fix $\eta=0.8$, and the mode number $N$=3, 4, 5, 6, respectively. In Fig. 6, we fix the mode number $N$=3, and four curves represent $\eta=0.2, 0.4, 0.5, 0.8$, respectively. Interestingly, in Fig. 3 and Fig. 6, all curves pass through the same point with $t=\frac{1}{2}$. For $N=3$, if $t=\frac{1}{2}$, we can get $P=\eta t^{2}(1-t)+(1-\eta)t^{3}=\eta*\frac{1}{8}+(1-\eta)*\frac{1}{8}=\frac{1}{8}$, and we can obtain $G=1$. Actually, if $t=1$, it is essentially the teleportation of the single-photon entanglement.

\section{Summary}
In summary, we presented  a NLA protocol for protecting multi-mode single-photon entanglement. With the help of VBSs and some
local single photons, the fidelity of the multi-mode single-photon entanglement can be increased with $t<\frac{1}{2}$. It is also shown that the $G$ does not change with
the number of spatial mode of the single-photon entanglement. We hope that this NLA protocol is useful in current quantum information processing.

\section{Acknowledgements}
The project is supported by the National Natural Science Foundation of China
(Grant Nos. 11104159, 11347110 and 61201164), and the Priority Academic Program Development of Jiangsu Higher Education Institutions.

\end{document}